
\normalspace
\overfullrule=0pt

\def\si{\sigma}
\def\am{{\alpha}^\mu}

\def\ss #1{\sum_{s} #1}
\def\e{\equiv}

\def\la{\lambda}

\def\bra #1{\left\langle#1\right\vert}
\def\ket #1{\left\vert#1\right\rangle}

\FRONTPAGE
\line{\hfill BROWN-HET-917}
\line{\hfill June -- July 1993}
\bigskip
\title{{\bf
SCATTERING OF DISCRETE STATES IN TWO DIMENSIONAL OPEN STRING FIELD
THEORY
     \foot{Work Supported in part by
the Department of Energy under
contract DE-FG02-91ER40688 -- Task A.}
     }}
\bigskip
\centerline{Branko Uro\v sevi\'c}
\centerline{\it Physics Department, Brown University, Providence, RI 02912,
USA}
\bigskip
\abstract
{This is the second in a series of papers devoted to open string field theory
 in two dimensions. In this paper we aim to clarify the origin and the role
 of discrete physical states in the theory. To this end, we study
 interactions of discrete states and generic tachyons. In particular, we
discuss at length four point amplitudes.
 We show that behavior of the correlation functions is governed by the
 number of generic tachyons involved and values of the kinematic
 invariants $s$, $t$ and $u$. Divergence of certain classes of
 correlators is shown to be the consequence of the fact certain
 kinematic invariants are non--positive integers in that case.
 Explicit examples are included. We check our results by standard
 conformal technique.}

\vfill
\endpage

\chapter{{\bf Introduction}}
This is the second in a series of papers devoted to open string
 field
theory (SFT) in two dimensions. In the previous paper, Ref. 1, we have
 constructed Witten's--like vertices for the open string field theory and
discussed
the quantization of the theory in Siegel's gauge ($b_0 = 0$). We
 have used,
then, the corresponding Feynman rules to calculate tachyon -- tachyon
scattering amplitudes. Our results were presented in the form of a sum over
 poles corresponding to tachyonic and (discrete) excited intermediate states.
Residues of these poles were
shown to match the results of Bershadsky and Kutasov, Ref. 2.
 For the summary of our notation and conventions reader may consult Ref. 1,
Sec.2.

It is well known that in addition to the massless tachyon, string theory in
 $2d$ also has discrete higher string states.
  The presence of discrete states (DS) in the spectrum of two
 dimensional strings, i.e. states which are physical only for
 some particular values of momenta, has been known for quite some time now.
 They were first discovered in the matrix model approach (Ref. 3,4), and
rediscovered in the Liouville approach (Ref. 5--11). In Ref 5, it
 was shown that BRST cohomology $H^{(*)}$ of the two dimensional string is
nontrivial for more than one ghost number. This is in sharp contrast with
 the 26-dimensional string theory, where only $H^{(1)}$ is
 nontrivial (Ref. 12). This novel feature of 2d string theory has its origin in
nontrivial background charge of the system.
 The question is then how the discrete states participate in the field theory.
For example, it was even suggested recently (Ref. 13)
 that in order to recover the results of Ref. 5 from
 the field theoretic point of view, one would need to modify
 Witten's classical action so as to include more than one ghost number.

 In this paper we would like to clarify the origin and the role of
 discrete states in the string field theory.  We first show (Sec. 2)
 that one can stay with the original Witten's action.
 The modification is redundant since the proper object to consider in
investigation of discrete states is a {\it gauge fixed} action.
 In particular, choosing Siegel's gauge is then equivalent to restricting
ourselves to the relative
 cohomology of the first quantized BRST operator.
To probe the dynamics of discerete states we study their scattering with
generic tachyons (T).
 The naive counting of the degrees of freedom for string excitatations in two
dimensions gives $2 - 2 = 0$, so
 that the excited (discrete) states form a set of measure zero with respect to
the functional integral measure. It
 is natural to expect, therefore, that the correlators involving these states
may not be well defined.
 Indeed, earlier calculations of the on -- shell scattering amplitudes have
indicated that DS -- DS amplitudes are intrinsically  divergent for the closed
(Ref. 14) as well as open strings (Ref. 15).
 It is then of certain importance to consider the calculation of the amplitudes
in SFT. In Sec. 5, we present a detailed analysis of the
 four--string scattering in the case when all four of the external states are
from $H^{(1)}$. Our expansion is found in the form of
 sum over poles. We have a simple criterion which allows us to determine
whether an amplitude is well--defined, diverges or vanishes.
 Namely, we consider the three kinematic invariants $s$,
 $t$, and $u$ (see Sec. 4). If some of them are integers we call the
corresponding channels {\it degenerate}.
 A degenerate channel can either diverge (if the kinematic invariant is
non--positive) or vanish (if it is positive).
 We provide examples which illustrate how this works. Clearly, less generic
tachyons we have -- more channels will degenerate.
 In fact, while $A_{T T T D}$ are always well--defined and contain an infinite
number of physical poles
 (which is the behavior one expects from string amplitudes),
 $A_{T T D D}$ amplitudes have more subtle behavior.
 Namely, there are three different subclasses of amplitudes of that type: $A_{T
T D D}^{+}$, $A_{T T D D}^{-}$
 and $A_{T T D D}^{deg}$.
 Subscript denotes the sign of the degenerate kinematic invariant. The
correlators of the
 first class are well--defined and have finite number of poles in two channels.
Amplitudes of the second
 class diverge and have infinitely--many poles in two channels. Class $A_{T T D
D}^{deg}$ is divergent and degenerates
 in all three channel. The same is true for the amplitudes of the type  $A_{D D
D D}$.

The paper is organized as follows. In Sec. 2  we show how discrete
 states originate from the field theoretic point of
 view. In Sec. 3 we discuss three point functions in
 the theory and show that they agree with the first quantized results. In Sec.
4, we consider four tachyon correlation
 function $A_{+ + - -}$. This is an example from which one can learn some
important lessons. In particular, we there encounter,
 for the first time, the degeneracy of an amplitude.
 The amplitude is shown to vanish. Sec.5 represents the main course of the
paper. There, we present a classification
 of the four point amplitudes with respect to their dynamical properties. We
show that divergence occures when,
 and only when, certain kinematic invariants are non--positive integers. In
particular, in Sec 5.1 we
 discuss $A_{T T T D}$ class of amplitudes.
 Sec. 5.2 is devoted to the analysis of the $A_{T T D D}$ class. We conclude
the section by proving
 that the amplitudes of the class $A_{D D D D}$ always diverge (Sec. 5.3). We
check our conclusions
 using the conformal technique. Finally, in Sec. 6 we present
 a brief overview of the main results
 and outline some of the open problems and possible directions of future work.

\chapter {{\bf Discrete States From String Field Theory}}

As mentioned in Introduction, the presence of DS in the physical spectrum
 of two dimensional strings is one of its most intriquing
 features. Their relevance was first discovered in context of matrix models
(Ref. 3, 4). In the continuum approach, discrete states are
 part of the spectrum which correspond to string excitations. They survive the
gauge fixing due to the nontrivial background charge.
 In general, BRST analysis is a particularly elegant tool for
 determining physical degrees of freedom in a covariant way. In the first
quantization, physical states are cohomology classes
 of the first quantized BRST operator $Q$. Unlike critical string theory, where
the only nontrivial cohomology group is $H^{(1)}$,
 two dimensional strings have physical states for more than one ghost number
(Ref. 5, 11). For the chiral sector, which corresponds to
 open strings, relative cohomology ($b_0 = 0$) of $Q$
 is nontrivial for $g = 0,\,  1$ and $2$. Below  we would like
 to discuss physical states from the field theoretic point of view.

Let us, first, summarize the notation and definitions introduced in
 Ref. 1. A string field is given by an arbitrary ket
 vector $\ket A = \ss \ket s  a_s \,$ (see Ref. 1, sec. 3). Kets
 $\ket s$ belong to a one--string Fock space $F$ and contain information about
string excitations. Coefficient functions $a_s$, on the other hand,
 depend solely on the center--of--mass coordinates.
 Second quantization elevates the coefficient functions to the dynamical level.

The open string field theory can be described by the Witten's action:

$$
W_{cl} = {1 \over 2} \int (A_1 \star QA_1 + {2 \over 3} A_1 \star A_1
\star A_1)
\eqno\eq
$$

\noindent
where the subscript '1' means that this field has the ghost number $g=1$. The
Witten's action is invariant
 under the gauge transformations ($g \, \Lambda \, =0$):

$$
\Delta A_1 = Q\Lambda + A_1 \star \Lambda - \Lambda \star A_1.
\eqno\eq
$$

\noindent
To prove this, one has to take into account the properties of string
 field multiplication and integration known as Witten's Axioms
 (Eq. I, $(3.3)$). In a nutshell, they state that string fields
 should be treated in analogy to differential forms, with first
 quantized BRST charge (cf. Eq. I, $(2.17)$) playing the role of
 the differential and the (first quantized) ghost number $g_s$
 providing for grading. It was shown in Ref. 12, that it is
 advantageous to introduce yet another grading -- $Z_2$
 Grassmann grading -- so that the coefficient functions $a_s$ are
 Grassman even or odd. Alternatively, one can define {\it target space}
 ghost number $G_s = 1 \, - g_s$ so that Grassmann parity of the
 coefficient functions is consistent with their ghost numbers: $(-)^{a_s} =
(-)^{G_s}$. Then, the {\it total parity}
 or, simply, parity of a string field is $(-)^{g_s} \, (-)^{a_s} = \, -1 $. We
demand that all dynamical string fields are overall {\it odd}.
 This enables us to present gauge fixed action in a very simple way.

Problem of gauge fixing for the systems with reducible gauge symmetry,
 such as string field theory, can be treated successfully by
 Batalin -- Vilkovisky (BV) formalism. Recent developments in this field
 provide us with a geometric understanding of BV formalism (Ref. 16)
 and are closely connected to the promise of a background
 independent formulation of the string field theory (Ref. 17,18).
 In that approach, one would like to postpone the fixing of gauge
 as much as possible. To formulate the perturbation theory, on the other hand,
we still need to choose one particular gauge. In SFT, two most popular
 choices are light--cone and covariant Siegel's gauge ($b_0 = 0$).
 In this paper we work in Siegel's gauge, although alternative
 gauges are tempting to be expored (see Ref. 13 as well as comments
 below). The gauge fixed action reads:

$$
W_{G.F.} = \, {1 \over 2} \, \int \, (\, A \star Q \, A \, + \, {2 \over
3} \, A \star A \star A \, -2 \, (b_0 \beta) \star A \, )
\eqno\eq
$$

\noindent
where string field $A$ contains all possible ghost numbers and field $\beta$
  is a Lagrange multiplier enforcing the gauge condition. It can be
 proven, along the same lines as in Ref. 12, that $W_{G.F.}$ satisfies
 the classical BV master equation, so that, at least on the tree level,
 it is a consistent gauge fixed action.

Concrete realisation of the Witten's operations  Eq. I, $(3.12-16)$, is
 subject to stringent constraints:

$$   \eqalign
{
&{}_{12 \dots n}{\bra V}\, (\sum_{r=1}^n {\am}^r_0 + Q^{\mu}) =
{}_{12 \dots n}{\bra
V} {\delta}^{(2)}\, (\sum_{r=1}^n p^{\mu r} + Q^{\mu})=0\, , \cr
&{}_{12 \dots n}{\bra V}\, (\sum_{r=1}^n {\si}^r_0 - 3) = {}_{12 \dots n}{\bra
V} {\delta}\,(\sum_{r=1}^n {\la}^r - \, 3)\, = \, 0. \cr
}
\eqno\eq
$$

\noindent
where $p^{\mu}$ is the momentum (matter and Liouville) of the string
 state, $\la$ is its ghost number and $Q^{\mu}$ is the background charge
 of the matter -- Liouville system (cf. Eq. I, $(2.2)$). First of
 these equations means that we are necessarily calculating the
 'bulk' amplitudes, i.e. that we are explicitly putting cosmological
 constant to zero. The second one represents the ghost number
 conservation. Nonvanishing amplitudes involve the external states which
 ghost numbers add up to $3$, or alternatively, which target space
 ghost numbers add up to zero. Another important property of the
 multi--string vertices is that they are cyclically
 symmetric: ${}_{12 \dots n}{\bra V} \, =
 {}_{n1 \dots n-1}{\bra V} = \, \dots =\,  {}_{23 \dots 1}{\bra V}$.

Now let us go back to the question of physical states. Naively, the
 problem consists of solving the free classical EOM $Q A_1 = \, 0$,
 modulo gauge transformations $\Delta A_1 = Q\Lambda$. It does
 reproduce part of the spectrum, namely $H^{(1)}$, but the rest
 of the states ($g=0$ and $g=2$) seem to be missing (classical
 field $A_1$ has ghost number $g=1$). To circumvent this problem,
 a modified classical action was proposed (Ref. 13) which includes
 more than one ghost number. We would like to argue that this is
 not necessary. In fact, instead of classical action, the  analysis
 should be based on its {\it gauge fixed} version:

$$
W_{lin} = \, {1 \over 2} \, \int \, (\, A \star Q \, A \,  -2 \, (b_0 \beta)
\star A \, )
\eqno\eq
$$

\noindent
Note that gauge fixed actions in the  Siegel's gauge ($b_0 = 0$) for both
  the original and the modified classical actions  have the
 same form $(2.5)$.
The action $(2.5)$ is invariant under the {\it target space}
 BRST transformations $s$:

$$   \eqalign
{
&s \, A_{\leq 1} \, = \, (Q A)_{\leq 1} \,  , \cr
&s \, A_{\geq 2} \, = \, (b_0 \beta)_{\geq 2}  \, , \cr
&s \beta \, = 0   \, , \cr
}
\eqno\eq
$$

\noindent
where the subscripts denote, as usual, ghost numbers $g$ of the
 states in question. One should bear in mind that $s$ acts on
 coefficient functions $a_s$ whereas $Q$ acts on states $\ket s$. From
 the relation $G= 1-g$ one inferes that target space BRST
 transformation $s$ increases the {\it target space} ghost number
 $G$ by one unit. The Eq. $(2.6)$ implies that, for $g \leq 1$,
 $sA_1 = Q A_0$, $sA_0 = Q A_{-1} \cdots$, whereas,
 for $g \geq 2$, one has that $sA_2 = b_0 \beta_{3}$,
 $sA_3 = b_0 \beta_{4} \, \cdots$. It is easy to check that
 $s$ is nilpotent off--shell.

Let us now prove that solutions of EOM corresponding to $(2.5)$, modulo
 BRST transformations $(2.6)$, exactly reproduce the physical
 spectrum of the theory.
The target space BRST symmetry is a residual symmetry left after
 the fixing of gauge. The EOM read:

$$
Q A_n  \, - b_0 \beta_{n+2} \, = 0 \, .
\eqno\eq
$$

\noindent
Substituting Eq. $(2.7)$ into Eq. $(2.6)$ one gets:

$$
s A_n =  \, Q A_{n-1}
\eqno\eq
$$

\noindent
for all $g = n$. It is now obvious that $s$ and $Q$ cohomology are exactly
 the same, so that $H^{(*)}$ is trivial except for $g=0$, $1$ and $2$.
 (The more accurate statement is that we have shown that the relative
cohomologies coincide. To consider the absolute cohomology, one would
 have to find a different gauge (Ref. 13) for which $b_0 \neq 0$.
In that case, there would be, also, physical states of $g=3$).

The appearence of states with ghost number $1$ ($g=1$) is predictable
 since they exist in higher dimensions as well. In two dimensions,
 $H^{(1)}$ consists of a massless tachyon (of generic momenta) and DS.
 Let us consider them in more details (cf Ref. 6). For a generic
 tachyon $T$ to be physical, mass--shell condition should be
satisfied ($Q = \, 2 \sqrt 2$ for $c=1$):

$$
{1 \over 2} k^2 \, - \, {1 \over 2} \, \beta \, (\beta + Q) = \, 1  \, ,
\eqno\eq
$$

\noindent
where $k$ is the matter and $\beta$ is the Liouville momentum. Solving
 for $\beta$, we see that tachyons may have two different Liouville
 dressings:

$$
k^{\mu}_{\pm} \, = \, (k, \, -{Q \over 2} \pm k) .
\eqno\eq
$$

\noindent
In an important special case, $k$ is an integer or half--integer multiple
 of $\sqrt 2$. Such a tachyon is called special or discrete tachyon.
 Discrete tachyons play very important role. Namely, they are the
 highest weights of the underlying $SU(2)$ symmetry of the spectrum.
 Starting from a discrete tachyon $V_{s,\, s}^{\pm}$ (here $s$ is a
non--negative integer or half--integer):

$$
V_{s,\, s}^{\pm} = \, c \, e^{i \sqrt 2\, s \, x}\, e^{(- \sqrt 2\, \pm \,
\sqrt 2 \, s )\, \varphi} \, ,
\eqno\eq
$$

\noindent
one can construct all of the $g = 1$ discrete states by applying the
 raising or lowering $SU(2)$ operators $H_{\pm} = {1 \over {2 \pi i}}
 \, \oint \, e^{\pm i \sqrt 2 \, x}$ to $(2.11)$ certain number
 of times:

$$
W_{s,\, n}^{\pm}\,  \propto \, (H_{-})^{s-n} \, V_{s, \, s}^{\pm}.
\eqno\eq
$$

\noindent
As we can see, discrete states also have positive or negative dressings,
 which correspond to positive or negative energies. In theories
 with non--vanishing cosmological constant, or in non--trivial background
 (black holes), sign of the energy is important. The non--zero
 cosmological introduces the Liouville wall at $- \infty$. This means
 that the wave functions corresponding to the negative energy states can
 not be normalized -- they are 'wrongly dressed'. In our case the
 cosmological constant is absent so, dynamically, there is no much
 difference between the two dressings. Even in that case, however,
 there is difference in the interpretation of the two branches of
 states (Ref. 6). In fact, positive branch of discrete states can be
 viewed as singular gauge transformations while there is no such
 interpretations for the negative branch. Neither of them are,
 of course, gauge artifacts and are physical degrees of freeedom.

Matter and Louiville fields enter the Eq. $(2.12)$ on different footing,
 since only the matter part contributs to excitations. In fact, it can
 be proved that such a gauge is a legitimate one. On the other hand,
 a gauge which would have only Liouville excitations is not.

For each DS of $g=1$ there are 'partners' of $g=0$ ('chiral ground
 ring') and $g=2$ (Ref. 9). These states play important role
 in spectrum generating symmetry of the theory ($W_{\infty}$) and
 their dynamics will be studied elsewhere. In what follows, we are
 interested in scattering of $g=1$ states only.

\chapter{\bf Feynman Rules and Three Point Functions}

In this section we summarize the Feynman rules and discuss the three
 point functions in the theory. After inserting coupling constant
 $g$ and integrating over the Lagrange multiplier field $\beta$,
 $(2.3)$ becomes:

$$
W_{g.f.} \, = {1 \over 2} \sum_{s,l} \, K_{s l}
\, a_{l} \, a_{s} +\, {g  \over 3} \, \sum_{s,l,m} V_{s l m} \, a_m \, a_l \,
 a_s \, (-)^{a_l}  \, .
\eqno\eq
$$

\noindent
where the kinetic term involves $K_{s l} \, \e \, {}_{21}{\bra V}
 {\ket s}_1 \, c_0  (L_0 - 1) \, {\ket l}_2$ and the interaction
 vertices are given by $V_{s l m} \, = \, {}_{321}{\bra V}
 {\ket s}_1 {\ket l}_2 {\ket m}_3$. Kinetic matrices $K_{s l}$ are
 invertible and their inverses, $D_{s l}$, are the free propagators
 of the corresponding cofficient fields.
Since in this work we are interested in tree amplitudes (the
 loop corrections are, undoubtedly, very interesting but more complicated
 since they necessarily involve closed strings, Ref. 12), free propagators
 give the two point functions. Three point functions are obtained
 in the standard fashion, using the Wick's Theorem:

$$
A_3 \,(s,l,m) = \, - g \, (-)^{a_l} \, V_{s l m} \, (1 + (-)^{R_s + R_l
 + R_m}),
\eqno\eq
$$

\noindent
where by $R$ we have denoted the {\it fermion number} of the
 corresponding state. It is an eigenvalue of the level operator which,
 acting on conformal and physical vacua, respectivelly, gives
 $R {\ket 0}\,  = \, {\ket 0}$, $R {\ket \Omega}\,  = \, 0$
 (cf. Ref. 1, Sec.2). In deriving $(3.2)$ we have used Grassmannian
 parity of the coefficient functions as well as cyclic symmetry of
 the vertices. This completes the derivation of the Feynman rules.

 Let us now consider the three point functions, Eq. $(3.2)$. Ghost
 number conservation allows for two different types of on--shell
 correlators: either all three particles have $g=1$, or each of them
 has a different ghost number (i.e.    $0$, $1$ and $2$). One notices
 that $g=1$ states are involved in both cases.
Consider in more detail the three--point functions where all three
 states are from $H^{(1)}$. There are two distinct classes of amplitudes
 of that type: $V_{T T D}$ and $V_ {D D D}$. Here, we have denoted a
 generic on--shell tachyon  by $T$ and all $g=1$ discrete states, including
 the discrete tachyons, by $D$. The fact that there is no vertices of
 the type $V_{T T T}$ or $V_{T D D}$ follows from the momentum
 conservation. Let us prove that the class $V_{T T T}$ is empty
 (the second claim is obvious). The momentum conservation gives:

$$  \eqalign
{
&k_1 \, + k_2 \, + k_3 = \, 0  \, , \cr
- &\sqrt 2 \, \pm \, k_1 \, \pm k_2 \, \pm k_3 = \, 0  \, , \cr
}
\eqno\eq
$$

\noindent
where $+$ ($-$) corresponds, as usual, to the positive (negative)
 chirality. Clearly, we can not take all signs in the second equation to
 be the same. So, we can express two of the momenta entering with the
 same sign in terms of the third one (using the first equation) and plug
 it back into the second equation. In this way the third momentum is
 completely determined -- it corresponds to a discrete tachyon. We
 have shown, therefore, that the amplitude does not belong to $V_{T T T}$
 but, rather, to $V_{T T D}$. It is customary to normalize the
 three -- tachyon amplitudes to be $1$.

Consider, now,  the most general scattering of one DS and two generic
 tachyons. In that case the momentum conservation gives:

$$    \eqalign
{
&\sqrt 2 \, n_1 +  k_2 + k_3  = \, 0   \cr
&\sqrt 2 \, (s_1 - 1) \pm k_2 \pm k_3 = \, 0 \cr
}
\eqno\eq
$$

\noindent
Here, the index '1' is reserved  for the discrete state, while '2' and
 '3' label the tachyons. Up to now, chiralities of the tachyons were
 arbitrary. However, for $k_i$ to be non--discrete, the determinant
 of the system $(3.4)$ must vanish. This means that both signs in $(3.4)$
 should be the same. Let us, for definiteness, take the '$+$' sign.
 Then, the compatibility dictates that $n_1 = s_1 - 1$. Simple example
 of the state of that type is $W_{{3 \over 2},\, {1 \over 2}}^{+} =
 \, (\, - (\partial x)^2 \, -  {i \over {\sqrt 2}} \,
 {\partial}^2 x\, ) \,  e^{i {\sqrt 2 \over 2}\, x}\,
 e^{{\sqrt 2 \over 2} \varphi}$. Calculation of the correlation function
 $A_3 = \langle W_{{3 \over 2},\, {1 \over 2}}^{+} \, T_{k_2}^{+}
 \, T_{k_3}^{+} \rangle$ performed by the standard conformal
 technique gives:

$$
A_3 = \, (k_2)^2 + \, {k_2 \over {\sqrt 2}}
\eqno\eq
$$

\noindent
To arrive at $(3.5)$ we have fixed, using the $SL(2, R)$ symmetry, the
 three points on the boundary to be $z_1 = 0$, $z_2 = 1$ and $z_3
 = \infty$. Also, we have used the momentum conservation $(3.4)$.
 Let us calculate the same amplitude from the field theoretic point
 of view. From $(3.2)$ one has that:

$$   \eqalign
{
A_3 &= \, - g \, (-)^{a_l} \, V_{s l m} \, (1 + (-)^{R_s + R_l + R_m})
 =  \cr
&= \, - 2 g  \, V_{W T T} = \, - 2 g \, (e^{-2 N_{00}} \,
 (N_{10}^{12})^2 \, (k_2 - k_3)^2 \, + \, N_{11}^{11} - \, N_{20}^{11} \, )=
\cr
&= - 2 g \, ( {27 \over 16} \, (\,  {4 \over 27} (4 (k_2)^2 +
 \, 2 \sqrt 2 k_2 + \, {1 \over 2}) \, - {5 \over 27} + \,
 {3 \over 27})=   \cr
&= - 2 g (\, (k_2)^2 + {1 \over {\sqrt 2}} k_2)  \, , \cr
}
\eqno\eq
$$

\noindent
where, in the third line, we have used the explicit expressions for the
 Neumann coefficients (see Ref. 19), and expressed, using the Eq.
 $(3.4)$, $k_3$ in terms of $k_2$. It is evident that, apart from
 an overall normalization factor, we have obtained the same
 amplitude. As explained above, the normalization is fixed by requiring
 that the three tachyon amplitudes are equal to unity, so that $-2 g = 1$.

In exactly the same way one can calculate $V_{D D D}$ amplitudes.
For $A_{+ + +}$, for example, the momentum conservation gives:

$$   \eqalign
{
&n_1 + n_2 + n_3 = \, 0 \cr
&s_1 + s_2 + s_3 = \, 1 \cr
}
\eqno\eq
$$

\noindent
and similary for the other possible chiralities. Using the associativity of
 the operator product expansion (OPE), a three point function can
 be represented as a linear combination of the two point functions.
 Coefficients in the expansion are proportional to the Clebsh - Gordon
coefficients, Ref. 6. To be nonzero, two point functions should pair
 the states with their reflected states: ${\ket A}^r_1 \, \e \, {}_{12}{\bra V}
{\ket A}^2  \,$. They are normalized to the momentum  delta function
 from which follows that the three point functions are equal to
 the OPE coefficients (Ref. 6). Such an amplitude is just a number
 (as oppose to an entire function in momenta, as it is the
 case for $V_{T T D}$). As an illustration, consider the same
 example as before, but instead of choosing both $+$ signs in
 the second of the equations $(3.4)$, let us take the second
 one to be $-$. Then, $k_2 = \, - {1 \over {\sqrt 2}}$, $k_3 = \, 0$
 and the amplitude vanishes, since: $ (k_2)^2 + {1 \over {\sqrt 2}}
 k_2 = \, 0$. This result holds, obviously, in both the first and
 the second quantized approaches.  Such an argreement between the
 two approaches clearly exists for all of the three point functions.

\chapter{{\bf An Instructive Example: Four Tachyon Scattering}}

As an introduction to our discussion of the four point scattering amplitudes,
 we would like in this section to consider an example which shows some
 of the peculiar properties of the two dimensional strings. Let us
 introduce two dimensional counterparts of the kinematic invariants
 in four dimensions: $s\, = \, {1 \over 2} (k_1 + k_2 + {Q \over 2})^2
 ,\, \,  t \, = \, {1 \over 2} (k_1 + k_3 + {Q \over 2})^2, \, \,
 u \, = \, {1 \over 2} (k_1 + k_4 + {Q \over 2})^2 $. It is
well
 known that for an arbitrary number of space--time dimensions
 $s +\,  t  +\,  u$ is an invariant quantity, determined solely by
 the
 masses of the external particles in question. In two dimensions, for
 four tachyons, one has that $s  +\,  t  +\,  u = \, 1$ (see $(5.22)$.
 Total amplitude is the sum  over $s$, $t$, and  $u$
 channels: $A_4^{(tot)} \, = \, A_4^{(s)}  + \, A_4^{(t)}  + \,  A_4^{(u)}$. As
a function of kinematic variables, an
 amplitude can be, generically, decomposed into the singular
 (which has simple poles in kinematic invariants) and the regular
 parts: $A = A_{sing} + \, A_{reg}$. It is important to note that only
$A_{sing}$ is physicaly relevant, since the physical information
 is contained in the residues of the poles. Two amplitudes agree
 if their singular parts agree. We will often be sloppy and suppress
the regular parts altogether.

For the $s$ channel, for example, one obtains (here $a \dots d$ stand
 for first \dots fourth external string states, respectively):

$$
A_4^{(s)}\, = \, -\, g^2 \, V_{b l a} \, (-)^a \, (1 + (-)^{R_a + R_l + R_b})
\, D_{m l} \, V_{d m c} \, (-)^d \, (1 + (-)^{R_c + R_m + R_d})
\eqno\eq
$$

\noindent
Summation over the repeated indices $l$, $m$ is implied in $(4.1)$.
 For the four tachyon scattering, one has the following
 kinematic-independent expression (cf. Ref. 1, Eq. $(5.5)$)

$$  \eqalign
{
&A^{(4)}_s \, \propto \, g^2 \, ({16 \over 27})^{{1 \over 2} (k_1 +
k_2 + {Q \over 2})^2}  \, ( {1 \over {{1 \over 2} (k_1 +
k_2 + {Q \over 2})^2}} \, + \,{{{4 \over 27}(k_1 - k_2) \cdot
 (k_3 - k_4)}
\over {{1 \over 2} (k_1 +
k_2 + {Q \over 2})^2 \, +1}}\,+\cr
& \cr
&+ \, {{{2 \over 81} (k_1 + k_2 + Q) \cdot (k_1 + k_2)
 \, +
{8 \over 729} ((k_1 - k_2) \cdot (k_3 - k_4))^2} \over
  {{1 \over 2} (k_1 +
k_2 + {Q \over 2})^2 \, +2}}\, + \cr
&+\, {{{124 \over 729} - \, {10 \over 729}
 ( (k_1 -
k_2)^2 \, + \,(k_3 -
k_4)^2 )} \over {{1 \over 2} (k_1 +
k_2 + {Q \over 2})^2 \, +2}}\,+ \cdots  ) \, \delta (\,
\sum_{i=1}^4 k_i\, +\, Q)
.  \cr
}
\eqno\eq
$$

 To obtain $t$ ($u$) channel contributions one is, simply, to
 substitute $2 \leftrightarrow 3$ ($2 \leftrightarrow 4$)
 in $(4.2)$.

In Ref. 1, we have analyzed the situation where three of the
 external tachyons are of one chirality and the fourth one is of
 the opposite. In particular, we have shown that the amplitude
 $(4.2)$ reproduces the Bershadsky--Kutasov amplitude (Ref. 2) in that
 kinematic region, if one compares the residues of the two
 expressions (see Ref. 1, Sec.5).

Let us, now, consider our main topic in this section, that is, a
 four--tachyon amplitude $A_{+ + - -}$. Bershadsky and Kutasov have
 argued that the total amplitude in that case should be zero. We would
 like to show how the same result appears from a field--theoretic
 point of view.

For that particular kinematics one has, for the matter momentum:
 $k_1 \, + \, k_2 = \, - (k_3 + k_4) = \, \sqrt 2$. It is easy to
 see that this implies that $s \, = \, 1$ and that, therefore,
 $t \, + u \, = 0$. This proves to be crucial for the vanishing
 of the amplitude. To see this, let us first calculate $t$ channel
 contribution. We obtain:

$$
A^{(t)}_4 \, = \, - g^2 \, ({16 \over 27})^t \,   ( \,
{1 \over t} \,  + {{{120 \over 729} \, + \,  {76 \over 729} \, t + \,
 {32 \over 720} \, (t \, + 1)^2} \over {t \, + 2}} \, + \, \cdots \, ).
\eqno\eq
$$

\noindent
Dots stand for higher level contributions. The $u$ channel amplitude
 is obtained from $(4.3)$ by $t \leftrightarrow u$. It is straightforward
 to check that almost all poles in $A^{(t)}_4$ ($A^{(u)}_4$) are
 {\it fake}, i.e., that they have zero residues. The only exception
 is the pole $t=0$ ($u=0$). Because of that, taken separately, singular
 parts of $A^{(t)}_4$ and $A^{(u)}_4$ do not vanish. This is not true
 for their sum, however, due to $t\, + u = \, 0$. In fact:
$$
A^{(t)}_4 \, + A^{(u)}_4 \, \propto -g^2 \, (({16 \over 27})^t \,
 {1 \over t} \, + \, ({16 \over 27})^u \, {1 \over u}) \, = \,
 -g^2 \, (\, ({16 \over 27})^t \, - \, ({16 \over 27})^{-t}\, )
 \, \, {1 \over t} \, .
\eqno\eq
$$

\noindent
As expected, the expression $(4.4)$ is regular at $t = \, 0$.
 Since $s = 1$, $A^{(s)}$ is a regular function as well. Thus, the
 singular part of $A^{(tot)}$ vanishes. We can add to $A^{(tot)}$
 an arbitrary entire function  without changing the physics. If
 we choose $A^{(reg)}$ to vanish, then $A^{(tot)} = 0$ as well.
 This completes the proof. In exactly the same way one can prove
 that $A_{+ - + -}$ vanishes. and that, more generally,
 $A_{n_1, m_1, \dots , n_k, m_k} = 0$, (where $n_i$ ($m_i$) is the
 number of consecutive $+$ ($-$)) for $k \geq 2$.

To summarize, there are couple of important messages from this simple
 example. First, to be able to draw physical conclusions, one is to
 consider all possible channels and not only one (as it is sometimes
 the habit). Second, physical information is contained in residues
 of the poles. Peculiarity of two dimensions is that, there, due
 to special kinematic restrictions, poles in kinematic variables
 can 'degenerate' -- instead of a variable we get an integer in the
 denominator. If this number is positive (as it was the case in
 this example) the amplitude becomes an entire function (or a number).
 If the number is non--positive (see below), one can anticipate
 the existence of an unbounded contibution to the sum. As we shall
 see below, this is the field--theoretic origin of the divergence
 of $2d$ amplitudes discovered in Ref. 14 in conformal approach.

\chapter{{\bf Four Point Amplitudes Involving Discrete States}}

In this section we analize four point amplitudes involving DS as well
 as tachyons. We are interested in correlation functions where all four
asymptotic states are from $H^{(1)}$. As in the three point case, four
 point amplitudes are severely restricted by the momentum conservation
 Law and can be classified in the similar fashion as it was done in
 Sec. 3 for the three pont functions. There are three different classes
 of correlators: $A _{T T T D}$, $A_{T T D D}$ and $A_{D D D D}$. Note that
 the four tachyon amplitude $A_{+ + + -}$ belongs to $A_{T T T D}$ rather
 than $A_{T T T T}$ since the negative chirality tachyon is fixed by
 the kinematics to be $W_{1 , -1}^{-}$. Trully belonging to $A_{T T T T}$
 class would be the amplitude considered in Sec. 4, $A_{+ + - -}$, but
 it vanishes. Clearly, there is no amplitudes of the type $A_{T D D D}$
 either. Before proceding to the detailed analysis of each one of
 the three classes, some general comments are in order.

A typical $s$ channel contribution to the four point
 amplitude is (see $(4.1)$):

$$
A_4^{(s)}\, \propto \,  \, -2 g^2 \,  \sum_l V_{b l a}
  \, D_{l} \, V_{d l c} \, =   \, \sum_{n \geq 0} \, {A_n\, (t)
 \over {s + \, n}}  \, ,
\eqno\eq
$$

\noindent
and similary for $t$ and $u$ channels. in previous section we have shown
 that  $s$ and, therefore, propagator $D_{l} \, \propto {1 \over
 {s + \, n}}$ can 'degenerate' and become a number (instead of
 a function of momenta) for some particular kinematics. In that
 case, $s$ was a positive integer ($s = \, 1$) and the amplitude
 was shown to vanish. Quite generally, if the amplitude degenerates
 in some channel, that channel either does not contribute or the
 amplitude diverges. This is determined by the sign of the
 degenerate kinematic invariant. To see this, one should bear in
 mind that $n \geq 0$, so that if $s$ is a positive number (in the
 degenerate case it is always an integer, cf. Sec. 5.3) amplitude $(5.1)$
  is an entire, bounded for finite $k$, function of $t$ and, as such,
 it is irrelevant. If $s \leq 0$, on the other hand, there is always
 an $n$ such that $s + \, n$ vanishes. This leads to the appearence
 of an unbounded term in the amplitude. This is the origin of
 divergencies in two dimensions.

One can arrive to the same conclusions using the conformal approach.
 Consider a generic correlator contribution $ \propto \, \int_0^1
 \, dx \, x^{k_1 \cdot k_2 \, + n_1} \, (1 - x)^{k_2 \cdot k_3 \,
 + n_2}$. Here, $n_1$ and $n_2$ are integers. Note that $s$ can be
rewritten as $s = {1 \over 2} \, (k_1 + {Q \over 2})^2 \, + {1 \over 2}
 \, (k_2 + {Q \over 2})^2 \, + \, k_1 \cdot k_2 \, + \, 1$.
 Since the first two summands are the masses of the corresponding
particles (integers in units of Regge slope) we see that $s$ and
 $k_1 \cdot k_2$ differ by, at most, an integer. Similary, $k_2
\cdot k_3$ ($k_1 \cdot k_3$) differ from $u$ ($t$) by, at most, an
 integer. In higher dimensions the amplitude is, in general, a
 meromorphic function of kinematic invariants  and is, therefore,
 well--defined. In two dimensions, when degeneracy occures, the
 exponents of $x$ and/or $1 - x$ can be negative integers. In that
 case the amplitude is clearly ill--defined. Value of the exponent, on
 the other hand, is determined by the value of the kinematic invariant
 in question. This proves the claim.

As an application, we check that the whole class of amplitudes $A_{T T T D}$
 is well--defined by simply showing that all three channels do not
 degenerate (Sec 5.1) . The opposite extreme are the amplitudes of
 the type $A_{D D D D}$ (see Sec. 5.3). They degenerate in all three
 channels. What is more, at least one of the kinematic invariants is
non--positive. Thus, the amplitudes of that class always diverge. In
 between the two extremes is the class $A_{T T D D}$ (Sec. 5.2).  The
correlators of that type have, at least, one degenerate channel. Some
 amplitudes from $A_{T T D D}$  are divergent while the others are
well--behaved. In what follows, we analyse in detail the dynamical
 properties of the four point amplitudes and present their classification.

\section{{\bf Amplitudes Involving One DS}}

Let us begin with $A_{T T T D}$ class. It contains, among the others, the
 four tachyon amplitude $A_{+ + + -}$. The properties of this amplitude
 are well--established (see I, Sec. 5). In this subsection, we
 would like to clarify some of the properties of $A_{T T T D}$ class as
 a whole. In particular, we prove that an arbitrary amplitude belonging
 to it is well--defined.

When a correlator involves an arbitrary discrete state $W_{s, n}^{\pm}$
  and three generic tachyons, the momentum conservation reads:

$$    \eqalign
{
&\sqrt 2 n_1 \, + k_2  + k_3 + k_4 = \, 0  \cr
-&2 \sqrt 2 \, \pm \sqrt 2 s_1 \pm k_2 \pm k_3 \pm k_4 = \, 0  \cr
}
\eqno\eq
$$

\noindent
If an amplitude is to belong to the class above, tachyons should be all
 of the same chirality and the following consistency condition must
 be valid:

$$
n_1 = \, \pm \, ( s_1 - 2) \, ,
\eqno\eq
$$

\noindent
Here, the upper (lower) sign corresponds to the positive (negative)
 chirality tachyons, respectively. It is easy to check that $s =
{1 \over 2} (k_1 + k_2 \, + {Q \over 2})^2 \, = \, \mp \, (\sqrt 2 k_2 \,
 + 2 n_1) \, - 1$ with the  similar expressions for $t$ and $u$. They
 are obtained from $s$ by substituting $k_2 \leftrightarrow k_3$
 ($k_2 \leftrightarrow k_4$). It is evident that none of the channels
degenerates. In accordance to the discussion above, this means that all
amplitudes of the type $A_{T T T D}$ are well--defined. Note that $\mp \,
 2 n_1 \, -1$ are {\it integers} so that the possible poles have the
 structure $k = \pm {N \over \sqrt 2}$, where $N$ is an integer.
 Dynamical properties of the $A_{T T T D}$ class can be summarized as
 follows: amplitudes of that class are meromorphic functions in discrete
momenta. There are no degerate channels in this case. Each amplitude
 of this class has an infinite number of proper physical poles.

As a simple example, take the discrete state to be $W_{{3 \over 2},
 \mp
 {1 \over 2}}^{+}$. Here, the upper--sign state is coupled
 to the positive--chirality tachyons and vice versa. One has:

$$
s = \, \mp \, \sqrt 2 k_2 \, ,\, \, t = \, \mp \, \sqrt 2 k_3 \, ,
 \, \, u = \, \mp \, \sqrt 2 k_4
\eqno\eq
$$

\noindent
Consider in more detail the first poles in the $s$ channel coressponding
 to the lower sign in Eq. $(5.4)$. The analysis of $t$ and $u$ channels
 goes along the same lines -- one is just to exchange the labeles as
 indicated above. The first potential pole is at $k_2 = 0$. The
 corresponding residue is:

$$
A_0 = \, 2 \, (N^{12}_{10})^2 \, (2 k_2 \, + {1 \over \sqrt 2})^2 \,
 - 2 \, N^{11}_{20} \, + \, 2 \, N^{11}_{11}  =  \, 0 \, ,
\eqno\eq
$$

\noindent
where we have used $k_2 = 0$, the momentum conservation, and the
 explicit expressions for the Neumann coefficients. Of course,
 the same conclusion can be reached without any calculations since
 the residue is proportional to the vertex containing $W_{{3 \over 2},
 {1 \over 2}}^{+}$ and the two discrete tachyons. Such a vertex, as it
 was shown in Sec. 3, vanishes. Next potential contribution is $n = 1$
($k_2 = \, - {1 \over \sqrt 2}$). In that case:

$$   \eqalign
{
A_1 = \, & 2 \,  ({16 \over 27})^{\sqrt 2 k_2 \, -1} (\,
 2 (N^{12}_{10})^2 N^{12}_{11} \, (k_3 - k_4) (2 k_2 \, +
{1 \over \sqrt 2}) \, +  \, \cr
&+ \, 2 \sqrt 2 \, (k_3 - k_4) (N^{12}_{10})^2 \, (\,
(N^{12}_{10})^2 \, (2 k_2 \, + {1 \over \sqrt 2})^2 \, +
 N^{11}_{11} \, - N^{11}_{20} \, )\, )  =  \cr
&= \, - 4 \sqrt 2 \, ({27 \over 16})^2 \, (N^{12}_{10})^2
 N^{12}_{11} \, \,  k_3 = \, - \, \sqrt 2  \, k_3  \, . \cr
}
\eqno\eq
$$

\noindent
The next one is the pole at $k_2 = \, - \sqrt 2$, with the residue:

$$
A_2 = \sqrt 2 \, k_3 - \, 2 (k_3)^2  \, .
\eqno\eq
$$

\noindent
In the same way we can calculate the higher orders. It is clear that
 the residues are entire functions in $k_3$ and that they are, in
 general, nonvanishig.

To check our conclusions, let us calculate the amplitude using the
 conformal technique. To this end, consider the correlation function:

$$   \eqalign
{
A &= \int_0^1 \, d x \, \langle W_{{3 \over 2}, {1 \over 2}}^{+} \,
 e^{i k_2 \cdot \phi}\,  c \, e^{i k_3 \cdot \phi} \, c \,
 e^{i k_4 \cdot \phi} \rangle =  \cr
&= \, \int_0^1 \, dx \,  x^{k_1 \cdot k_2} \, (1 - x)^{k_2 \cdot k_3}
 \, (\, {{k_2 \over \sqrt 2} \, + (k_2)^2 \over x^2} \, +  2
 { k_2 k_3  \over x} \, +  {k_3 \over \sqrt 2} \, +
 (k_3)^2 \, ) \cr
}
\eqno\eq
$$

\noindent
In passing from the first to the second line in Eq. (4.8) we have
 used the Wick's Theorem, fixed the $SL(2,R)$ gauge by choosing
 $z_1 = 0$, $z_3 = 1$ and $z_4 = \infty$, and integrated over the
 $z_2 = x$ ($0 \leq x \, \leq 1$). We have used, also, the on--shell
 conditions. For the kinematics in question $k_1 \cdot k_2 = 1 \, +
 \sqrt 2 k_2$ and $k_2 \cdot k_3 = \, - 2 \, - \sqrt 2 (k_2 + k_3)$.
 It is straightforward to see that the amplitude is well defined on
 the entire complex plane $k_2$, excluding the discrete set of points
 $\sqrt 2 \, k_2 \, + N = \, 0$, where it has the simple poles. The
 explicit expression can be easily found:

$$  \eqalign
{
A = &\, (\, {k_2 \over \sqrt 2} \, + (k_2)^2\, ) {\Gamma (\,\sqrt 2 k_2)
 \Gamma (k_2 \cdot k_3 \, +1) \over \Gamma ( \sqrt 2 k_2  \, + k_2
\cdot k_3 \, +1)} \, + \, (\, {k_3 \over \sqrt 2} \, + (k_3)^2\, )
 \times \cr
&\times \, {\Gamma (\,\sqrt 2 k_2 \, +2) \Gamma (k_2 \cdot k_3 \, +1)
 \over \Gamma ( \sqrt 2 k_2  \, + k_2 \cdot k_3 \, +3)}\,
+ \, 2 k_2 \, k_3 \, {\Gamma (\,\sqrt 2 k_2 \, +1)
 \Gamma (k_2 \cdot k_3 \, +1) \over
 \Gamma ( \sqrt 2 k_2  \, + k_2 \cdot k_3 \, +2)}\, .  \cr
}
\eqno\eq
$$

\noindent
Positions of the poles are the same as predicted from SFT. Let us compare
 the residues. To calculate them we use the well--known relation
 $\Gamma (z) = \, {\Gamma (z+1) \over z}$.  The residue corresponding
 to $k_2 = 0$ vanishes. Note that only the first summand in $(5.9)$
 has a potential pole for that value of $k_2$ since $\Gamma
 (\sqrt 2 \, k_2) \propto {1 \over \sqrt 2 \, k_2}$,  but it is killed
 by the factor $k_2 \, (k_2 + \, {1 \over \sqrt 2})$ which multiplies it.
 The next residue corresponds to $k_2 = \, - {1 \over \sqrt 2}$. In that
 case, the potential contribution from the first summand is, again,
 suppressed by the prefactor. However, the third summand contributes
 to the residue which reads: $2 \, k_2 \,  k_3 = \, -  \,  \sqrt 2 \,
  k_3$. Starting from $n \geq 2$ all three summands in $(5.9)$ begin
contributing to the residues, which are polinomials in $k_3$. Again, one
 readily checks that the values of the residues calculated from
 the amplitude $(5.9)$ match the ones calculated from SFT.

It is rather amusing to observe somewhat special role of the
 $W_{{3 \over 2}, \pm {1 \over 2}}^{+}$ states. Namely, these are the
 only two states compatible {\it simultaneosly} with $V_{T T D}$ and
 $A_{T T T D}$. In fact, the compatibility conditions are:

$$  \eqalign
{
&n = \, \pm \, (s - 1) \, , \cr
&n = \, \pm \, (2 - s) \, .\cr
}
\eqno\eq
$$

\noindent
Each of the two systems of equations has a solution, namely $s = \,
 {3 \over 2}$, $n = \, \pm \, {1 \over 2}$. These values correspond to
 the abovementioned states. It is unclear, however, whether this
 peculiar property of the $W_{{3 \over 2}, \pm  {1 \over 2}}^{+}$ states
 has some deeper physical meaning.

\section{{\bf Amplitudes Involving Two DS}}

Let us focus, now, on the properties of $A_{T T D D}$ class. We clarify,
 first, the conditions under which an amplitude belongs to that class.
 If we are given two arbitrary DS and two generic tachyons, the
 conservation of momentum tells us that:

$$   \eqalign
{
&\sqrt 2 \, n_1 + k_2 \, + k_3 \, + \sqrt 2 n_4 = \, 0   , \cr
-&2 \sqrt 2 \, \pm \, \sqrt 2 \,  s_1 \, \pm  \, k_2 \, \pm k_3
 \, \pm \, \sqrt 2 \, s_4 = \, 0   \, , \cr
}
\eqno\eq
$$

\noindent
where '1' and '4' label the discrete states and '2' and '3' the
 generic tachyons. It is clear that the tachyons should be of the
 same chirality.
 For concreteness, let us take it to be negative. Then, the consistency
 requires that:

$$
n_1 \, + n_4 = \, 2 \, \mp s_1 \, \mp s_4
\eqno\eq
$$

\noindent
Since the momenta $k_1$ and $k_4$ are fixed (discrete), the
 amplitude degenerates in, at least, one channel -- $u$ channel.
 Whether
 or not it degenerates in the other two can be easily determined.
 Since:

$$   \eqalign
{
&k_1 \cdot k_2 = \, \sqrt 2 \, (n_1  \,  \pm s_1 \, - 1) \, k_2
\, + \, 2 \, (\, \pm s_1 - 1)  \, ,  \cr
&k_1 \cdot k_3 = \, \sqrt 2 \, (n_1  \,  \pm s_1 \, - 1) \, k_3
 \, + \, 2 \, (\, \pm s_1 - 1)  \, ,  \cr
&k_2 \cdot k_3 = \, 2 \, (n_1 + n_4 - 1)   \, , \cr
}
\eqno\eq
$$

\noindent
$s$ and $t$ channels (simultaneously) degenerate if and only
 if $n_1 \, +  s_1  \, - 1 = \, 0  = \, n_4 \, +  s_4  \, - 1 = \, 0$.
 In that case, $k_1 \cdot k_2 =  \, k_1 \cdot k_3 = \,  2 \,  (s_1 - 1)$.
We refer to such an amplitude as {\it totally degenerate}. To determine
 the
 DS involved in a totally degenerate  correlator we use the fact that
 $- s \, \leq n \, \leq \, s$. It is easy to see that both discrete
 states should be taken to be $W_{{1 \over 2} \, , {1 \over 2}}^{+}$.
 Then, $k_1 \cdot k_2 = \, - 1$ and $k_2 \cdot k_3 = \, 2 \,
 (n_1 \, + n_4 - 1) = \, 0$, so the amplitude reads:

$$
A_{deg} = \,  \int_0^1 \, {d x \over x}
\eqno\eq
$$

\noindent
The integral on the right hand side is, evidently, ill--defined. The same
 result can be conjectured utilizing the SFT result for the scattering of
 four tachyons, Eq. $(4.2)$.  The first term in the expansion for
 $A_{deg}^{s}$ is ${ 1 \over s} \, $ and, since $s = 0$, it diverges.
Channel $t$ behaves in the same way. Since $u = 1$, channel $u$,
 on the other hand, does not contain any unbounded
summands. Divergence of the integral $(5.14)$, therefore, shows up
 in the field -- theoretic approach through the presence of unbounded
 summands in the amplitude.

Degeneracy in all three channels of $A_{T T D D}$ is not, however,
 the typical property of that class.
Much more common is the situation where only
one channel (say, $u$) denerates, while the other two ($s$ and $t$) do
 not. In that case, as explained above, dynamics is determined by the
 sign of $u$, or, equivalently, by the value of the product $k_2
 \cdot k_3 = \, 2 \, (n_1 + n_4  - 1) = \, u - 1$.
{}From the conformal field theory point of view, that sign determines
 the convergence properties on the upper limit of the Koba -- Nielsen
 integral. One can, thus, subdivide the correlators $A_{T T D D}$ into
 the three subclasses: $A_{T T D D}^{deg}$, $A_{T T D D}^{+}$ and $A_{T T D
D}^{-}$. The totally degenerate class $A_{T T D D}^{deg}$ has been
 discussed above. Let us, therefore, focus on the other two.

Amplitudes $A_{T T D D}^{\pm}$ contain physical intermediate
 particles. Note that $s$ channel poles, for example, originate
 from the expression $s \, + n$ in the denominator (see Eq. $(5.1)$).
 We have:

$$   \eqalign
{
s &= \, {1 \over 2} \, (k_1 + \, k_2)^2 \, = \, {1 \over 2} \,
 (\sqrt 2 \, n_1
+ \, k_2)^2 \, - \, {1 \over 2} \, (\sqrt 2 \, (1 \, \mp s_1) \,
 + k_2)^2 = \, \cr
&= ( n_1 - (1 \, \mp s_1) ) \, ( \sqrt 2\, k_2 \, +
(n_1 \, + (1 \, \mp s_1) ) \, ,  \, \cr
}
\eqno\eq
$$

\noindent
and although $n_1$ and $s_1$ can be half--integers, their linear
 combinations $n_1 - (1 \, \mp s_1)$ and $n_1 + (1 \, \mp s_1)$ are
 always
{\it integers}. The main difference between the two classes stems from
 the fact that, for $A_{T T D D}^{+}$, degenerate channel is a bounded,
 entire function ($u \geq 1$), while $A_{T T D D}^{-}$ has an
 unbounded contribution ($u \leq 0$).
Thus, $A_{T T D D}^{+}$ correlators are well--defined while
 $A_{T T D D}^{-}$ are not.

To see this in more detail, consider first the $A_{T T D D}^{+}$ class.
 A generic contribution is of the form:

$$
A  = \, \int_0^1 \, x^a  \,  (1 - x)^n \, d x
\eqno\eq
$$

\noindent
where $a$ is a variable and $n$ is (non--negative) integer constant.
 It is well--known that the integral of the type $\int_0^1 \, d x \,
 x^{a - 1}$ can be analitically continued, for all $a \neq 0$, to
 $\int_0^1 \, d x \, x^{a - 1} = {1 \over a}$.
The integral $(5.16)$ is of the that type (one is just to use Newton's
 binomial expansion formula). So, amplitudes of the class
 $A_{T T D D}^{+}$
 are well--defined and have finite number of poles in two
 differrent
 channels.

For example, let us take the two discerete states to be $W_{{3 \over 2} \,
 ,{1 \over 2}}^{+}$ and $W_{{1 \over 2}\, , {1 \over 2}}^{-}$. In that
 case, the correlation function $\langle W_{{3 \over 2} \,
 , {1 \over 2}}^{+}
 \, T_{k_2}^{-} \, T_{k_3}^{-} \,  W_{{1 \over 2} \, , {1 \over 2}}^{-}
 \rangle$ is:

$$
\langle W_{{3 \over 2} \, , {1 \over 2}}^{+} \, T_{k_2}^{-}
 \, T_{k_3}^{-} \,  W_{{1 \over 2} \, , {1 \over 2}}^{-} \rangle = \,
\int_0^1 \, d x \, x^{\sqrt 2  k_2\,  + 1} \, (\,
{{k_2 \over \sqrt 2} \, + (k_2)^2 \over x^2} \, +
  2 { k_2 k_3  \over x}
\, +  {k_3 \over \sqrt 2} \, + (k_3)^2 \, )
\eqno\eq
$$

\noindent
where we have used the fact that $k_1 \cdot k_2 = \, \sqrt 2 \, k_2 +
 1$, $k_2 \cdot k_3 \, = \, 0$ or, alternatively, that
 $s = \, \sqrt 2 \,
 k_2$, $u = 1$ in that kinematics.
  The amplitude $(5.17)$
 becomes:

$$
A^{s} \propto \, { 1 \over \sqrt 2 \, k_2 + 1}
\eqno\eq
$$

\noindent
Here, as usual, we have left out the finite part of the
amplitude.

The same result can be obtained from the field theoretic point of view.
 In fact, one can use $(5.5-5.7)$ to show that, in that case, $A_1 = \, 1$
 and $A_{n \neq 1} = \, 0$.
In exactly the same way we find that the $t$ channel amplitude is
 $A^{t} = \,
 {1 \over \sqrt 2 \, k_3 \, + 1}$ and that the $u$ channel
 contribution
 is of the form: $\sum_{n=0} \, {A_n \over n \, + 1}$, where $A_0 =
 \, 0$,
$A_1 = \, {27 \over 16} \, {1 \over 2} \,  (\, 8 \, (N_{10}^{12})^4 \,
 + 3 \, (N_{10}^{12})^2 \, N_{11}^{12}\, )\, \sqrt 2\, (k_3 - k_2) = \,
 {10 \, \sqrt 2 \over 27} \, (k_3 - k_2)$ and so on. We see that $A^{u}$
 is an entire, bounded function of momenta.
The amplitude has {\it finite} number (one in each channel) of poles
 in two different channels. One can readily see that the same holds
 for any $A_{T T D D}^{+}$ correlator.

Quite different is the situation when $A_{T T D D}^{-}$ is considered.
 In that case $u \leq 0$ and the amplitudes have the following form:

$$
A  = \, \int_0^1 \, {x^a \over (1 - x)^n} \, d x
\eqno\eq
$$

\noindent
for a positive integer $n$. Such an integral clearly diverges on the upper
 limit ($x = 1$). To illustrate the situation, consider the correlation
 function $\langle W_{1 \, , 1}^{+} \, T_{k_2}^{-} \, T_{k_3}^{-} \,
  W_{{1 \over 2} \, , -{1 \over 2}}^{+} \rangle = \, \int_0^1 \,
 {x^{\sqrt 2 \, k_2} \over 1 -x} \, d x$. This amplitude is of the type
 $(5.19)$ with $n = 1$ and $a = \, \sqrt 2\, k_2$. To make some sense out
 of that expression, let us perform a simple trick. It consists of
 expanding the ${1 \over 1 - x}$ in power series and formally integrating
 term by term. We say 'formally' since the geometric series diverges
 for $x = 1$, so that we do not have, strictly speaking, the right to
 exchange the order of summation and integration. Nevertheless, let us
 do just that. Then, the amplitude  yieds:

$$  \eqalign
{
&\langle W_{1 \, , 1}^{+} \, T_{k_2}^{-} \, T_{k_3}^{-} \,
 W_{{1 \over 2} \, , -{1 \over 2}}^{+} \rangle = \, \int_0^1 \,
 {x^{\sqrt 2 \, k_2} \over 1 -x} \, d x  =  \, \cr
&\sum_{n = 0}^{\infty} \, {1 \over \sqrt 2 \, k_2 \, + n \, + 1} = \,
 \sum_{n = 0}^{\infty} \, {1 \over s \, + n }  \, \cr
}
\eqno\eq
$$

\noindent
where, in the last step, we have used the fact that $s = \sqrt 2 \, k_2 + 1$.
 It is tempting to check the validity of this formula
 employing the SFT approach. Since the example in question is nothing
 but the scattering of
 four tachyons (some of them of the discrete momenta), one can, once again,
 use the general expression for the four tachyon amplitude $(4.2)$.
 Clearly, poles have the same position in both approaches. The real question
 is whether they have the same residues. From $(5.20)$ we see that the
 amplitude has unusual property -- all residues are equal to unity. Let
 us show that this is the case from the field theory approach. In
 fact, the first three residues are:

$$   \eqalign
{
&A_0 = ({16 \over 27})^0 = \, 1  \, ,  \cr
&A_1 = ({16 \over 27})^{- 1} \, {4 \over 27} \, (k_1 - k_2)
 \cdot (k_3 - k_4) =
\, {27 \over 16} \, \times \, {4 \over 27} \, \times \, 4 \,
  = \, 1,  \cr
&A_2 = \, ({27 \over 16})^2 \, \times \, ( - \, {4 \over 81} \,
 + \, {36 \times 8 \over 729} \, +
{124 \over 729} \, - {12 \times 10 \over 729}) = \, 1 \, ,  \cr
}
\eqno\eq
$$

\noindent
where, in the last line, we have presented, as separate summands,
 the numerical values of the four terms contributing to the residue at
 $s = \,  - 2$ in $(4.2)$. Although by now reader should have been
 convinced that the two approaches give the same results, it is nice
 to see how all these messy coeficients add up to $1$. We see that such
 an amplitude has infinitelly many physical poles. Their residues are
 numbers. The same properties has $t$ channel. On the other hand, $u = 0$
 so that $u$ channel degenerates and contains an unbounded contribution,
 ${1 \over u}$. This is where one can trace the 'bad' behavior of the
 integral. Thus, although the total amplitude is ill--defined it has a
well--defined 'piece' from which one can draw physical conclusions.
 'Changing the order of summation and integration' is, thus, the operation
 which {\it removes} the divergence from the pole expansion. One can clearly
generalize the above discussion to any amplitude of the $A_{T T D D}^{-}$
 class. Therefore, $A_{T T D D}^{-}$ amplitudes have infinitely many
 physical poles in two different channels and are divergent in the third.

\section{{\bf The Amplitudes Involving Four DS}}

To finish up this section, let us briefly comment on the third class of
correlators, $A_{D D D D}$. In that case, all three channels degenerate,
 and the amplitudes are ill -- defined. To prove it, it is enough to show
 that for every amplitude of that class there is at least one non--positive
kinematic invariant. This follows from the fact that:

$$
s + \, t + \, u = \, \sum_{i = 1}^4 \, {1 \over 2} \,
 (k_i + {Q \over 2})^2 \, \, + 1 = \, 1 - \, \sum_{i = 1}^4 \, R_i
\eqno\eq
$$

\noindent
where $R_i$ are the levels of the four asymptotic states  (see Sec 3.).
 Since $R_i \geq  \, 0$, and the equality holds only for tachyons,
 we have that:

$$
s + t + u \, \leq 1 \, .
\eqno\eq
$$

If we can show that the kinematic invariants are integers then we are,
obviously, done. Let us, for example, show that $s$ is an integer
 (the other two cases are treated similarily). We know that $s$ can be
represented as $s = {1 \over 2} \,
(k_1 + {Q \over 2})^2 \, + {1 \over 2} \, (k_2 + {Q \over 2})^2 \, + \, k_1
\cdot k_2 \, + \, 1$ . Since the masses of the external particles
 are integers, it is enough to show that $k_1 \cdot k_2$ is an integer.
 If we denote by $\sqrt 2 \, n_i$, $- \, \sqrt 2 \, \pm \, \sqrt 2 \, s_i\,
 $ the matter and Liouville momenta of the $i$--th string, the product can be
rewritten as: $k_1 \cdot k_2 = \, 2 \, n_1 n_2 \, - \, 2 s_1 s_2 \, - \,
 (2 \, \mp 2 \, s_1 \, \mp 2  \, s_2)$. The term in parentheses is
 clearly an integer (since $n_i$, $s_i$ are integers or half--integers).
 Also, although  $2 \, n_1 n_2$ and $ 2 s_1 s_2$, taken separately,
 may be half--integers, their difference is always an integer. Thus,
 we have proven that, whenever degenerate, kinematic invariants are
 {\it integer -- valued}. By the same token, using $(5.23)$, we
 have proven that at least one of the kinematic invariants is
 non--positive. Since every amplitude of the class $A_{D D D D}$ has
 at least one divergent channel, they are all ill--defined.

 \chapter{{\bf Concluding Remarks}}

In the present work we have aimed to better understand the dynamics of
 discrete states in open string field theory. In particular, we have shown
 that the origin of divergencies in $2d$ is rather simple to understand.
 Namely, they stem from the fact that, for certain kinematics, kinematic
invariants become non--positive integers. In that case, the amplitude,
 presented as sum over poles, has an unbounded contribution tantamount
 to the divergence of the corresponding Koba--Nielsen integral. We
 have seen, also, that, from the dynamical point of view, discrete
 tachyons are not at all different from the other discrete states.
 They differ dramatically, on the other hand, from the generic tachyons
. As far as convergence of amplitudes is concerned one can state a simple
 thumb rule: more generic tachyons -- better the convergence.

There are several important questions which our discussion left
 open or partially unanswered. First, our formalism is adapted only
 for calculations of the bulk amplitudes, i.e. cosmological constant is
 absent in this approach. Clearly, this makes a difference since presence
 of the cosmological constant leads to additional divergences, due
 to charge screening integrals (Ref. 14). Knowledge of bulk amplitudes
 allows one, in principle, to deduce non--bulk correlation functions using
 the method outlined in (Ref. 7). However, it is clearly adventageous to
calculate them directily. To do that, one would have to adapt the
 formalism of Ref. 1 so as to include nontrivial cosmological constant.
 This is an important problem which still needs to be solved.
We have, at present, concentrated on the correlation functions
 involving states from $H^{(1)}$ only. This is, certainly, very important
 class of amplitudes since $H^{(1)}$ contains, among the rest,
 generic tachyons. However, one would definitely like to know how the
 behavior of the correlation functions changes if some or all of
 the discrete states are from $H^{(0)}$ and $H^{(2)}$ (these are the
 other two relative cohomoly classes, see Sec. 3). The work on this
 subject is in progress and will be reported eslewhere. The related
 problem is the the role which $W_{\infty}$ symmetry plays in
 string field theory.

It is important to better understand the question of gauge fixing in
 $2d$ SFT. As we have shown in Sec. 3, choosing the Siegel's gauge is
 equivalent to restricting the physical spectrum of open strings to
 their relative cohomology. In Ref. 13, search for different gauges,
 which would allow ghost number three states, is advocated. How to
 choose a gauge fixed action is not just an academic problem since
 this is {\it the} action upon which Feynman rules are constructed.
 While Siegel's gauge is certainly consistent and gives correct tree
 results,  the issue is far from beeing completelly understood. Related
 to this is the problem of constructing an effective tachyonic field
 theory ('collective field theory') starting from the Witten's gauge
 invariant formulation.

The last, but not least, is the problem of going off--shell. Indeed,
 we have seen that scattering amplitudes of the class $A_{D D D D}$
 diverge. The same is true for $A_{D D T T}^{deg}$ and $A_{D D T T}^{-}$.
 To make these amplitudes sensible, one may have to regularize them.
 In the framework of the first quantization any $SL(2,R)$ invariant
regularization will do (Ref. 14) -- there is no physical reason of
 choosing one instead of the other. In field theory, on the other hand,
divergence is the consequence of degeneracy which, in turn, is the
 consequence of a peculiar two dimensional kinematics. One way to
 remove the degeneracy and, therefore, divergence is to go off -- shell.
 This is not, however, a trivial problem since while it is intuitivelly
 fairly clear what the off--shell generic tachyon is, it is not quite so
 for the states defined, at least on--shell, only for some particular values
 of momenta. The issue of off--shell amplitudes has been discussed in
 Ref. I, 28 but it, definitely, needs more attention.

\noindent
{\bf Acknowledgments.} I am grateful to A. Jevicki for his
guidance and support throughout the work on this project
 and to M. Li for making valuable comments on the manuscript.

\vfill
\endpage

\centerline{\it REFERENCES}
\bigskip

\pointbegin
B. Uro\v sevi\'c, {\it Phys. Rev. D15} {\bf 47} (1993) 5460;

\point
M. Bershadsky and D. Kutasov, {\it Phys. Lett.} {\bf B274} (1992) 331;
PUPT-1315, HUPT-92/A016.

\point
I.R. Klebanov, 'String theory in two dimensions', in 'String Theory and
 Quantum Gravity', Proceedings of the Trieste Spring School 1991,
 eds. J. Harvey et al., (World Scientific, Singapore, 1992).

\point
A. Jevicki, 'Developments in $2d$ String Theory', Lectures given at the
 Spring School on String Theory, Trieste, April 1993, Brown HET--918.

\point
B. Lian and G. Zuckerman, {\it Phys. Lett.} {\bf B254} (1991) 417;
{\it Phys. Lett.} {\bf B266} (1991) 2.

\point
A. M. Polyakov, {\it Mod. Phys. Lett.} {\bf A6} (1991) 635;
Preprint PUPT (Princeton) -1289 (Lectures given at 1991 Jerusalem
Winter School).

\point
D. Kutasov, "Some Properties of (non) Critical Strings",
PUPT-1272, 1991.

\point
J. Avan and A. Jevicki, {\it Phys. Lett.} {\bf B266} (1991) 35;
{\bf B272} (1992) 17.

\point
E. Witten, {\it Nucl. Phys.} {\bf B373} (1992) 187;
I. Klebanov and A. M. Polyakov, {\it Mod. Phys. Lett} {\bf A6}
(1991) 3273;
N. Sakai and Y. Tanii, {\it Prog. Theor. Phys.} {\bf 86} (1991) 547;
Y. Matsumura, N. Sakai and Y. Tanii, TIT (Tokyo) -HEEP 127, 186 (1992).

\point
E. Witten, B. Zwiebach, {\it Nucl.Phys.} {\bf B377} (1992) 55.

\point
P. Bouwknegt, J. Mc. Carthy and K. Pilch, Comm. in Math. Phys. {\bf
145} (1992) 541; CERN-TH.6279/91 (1991); K. Itoh, N. Ohta,
 Report No.
{\bf OS-GE-22-91} September 1991.

\point
C. Thorn, {\it Phys. Rep.} {\bf 174} (1989) 1.

\point
N. Sakai and Y. Tanii, {\it Mod. Phys. Lett.} {\bf A7}
(1992) 3486.

\point
M. Li, {\it Nucl. Phys.} {\bf B382} (1992) 242;

\point
I. Ya. Aref'eva, P. B. Medvedev and A. P. Zubarev,
Interaction of $d=2$ $c=1$ Discrete
States from String Field Theory (preprint SMI).

\point
A. Schwarz, `Geometry of Batalin--Vilkovisky quantization`,
 UC Davis preprint,  July 1992.

\point
E. Witten, 'On background independent open--string field theory`.

\point
H. Hata and B. Zwiebach, 'Developing the covariant
 Batalin--Vilkovisky
approach to string theory',
MIT-CTP-2184, to appear in Annals of Physics.

\point
S. Samuel, {\it Nucl.Phys.} {\bf B296} (1988) 187.

\endpage

\end